\def\ra{\rangle}
\def\la{\langle}
\def\be{\begin{equation}}
\def\ee{\end{equation}}
\def\ba{\begin{array}}
\def\ea{\end{array}}

\def\bf{{\textbf}}

\documentclass[showpacs,twocolumn,amsmath]{revtex4}
\usepackage{epsfig}
\usepackage[OT2,OT1]{fontenc}
\def\qed{\leavevmode\unskip\penalty9999 \hbox{}\nobreak\hfill
     \quad\hbox{\leavevmode  \hbox to.77778em{%
               \hfil\vrule   \vbox to.675em%
               {\hrule width.6em\vfil\hrule}\vrule\hfil}}
     \par\vskip3pt}

\newtheorem{theorem}{Theorem}

\begin{document}
\title{Characterizing multipartite entanglement by violation of CHSH inequalities}
\author{Ming Li$^{1,5}$}
\author{Huihui Qin$^{2}$}
\author{Chengjie Zhang$^{3}$}
\author{Shuqian Shen$^{1}$}
\author{Shao-Ming Fei$^{4,5}$}
\author{Heng Fan$^{6}$}

\affiliation{ $^{1}$ School of Science,
China University of Petroleum, Qingdao 266580,  China\\
$^2$ Beijing Computational Science Research Center, Beijing 100193, China\\
$^3$ School of Physical Science and Technology, Soochow University, Suzhou, 215006, China\\
$^4$ School of Mathematical Sciences, Capital Normal University,
Beijing 100048, China\\
$^5$ Max-Planck-Institute for Mathematics in the Sciences, Leipzig
04103, Germany\\
$^6$ Beijing National Laboratory for Condensed Matter Physics,
Institute of Physics, Chinese Academy of Sciences, Beijing 100190,
China}

\begin{abstract}
Entanglement of high-dimensional and multipartite quantum systems
offer promising perspectives in quantum information processing. However, the characterization and measure
of such kind of entanglement is of great challenge.
Here we consider the overlaps between the maximal quantum mean values and the classical bound of the CHSH
inequalities for pairwise-qubit states in two-dimensional subspaces.
We show that the concurrence of a pure state in any high-dimensional multipartite system can be equivalently
represented by these overlaps. Here we consider the projections of
 an arbitrary high-dimensional multipartite state to two-qubit states. We investigate the non-localities of these
 projected two-qubit sub-states by their violations of CHSH
  inequalities. From these violations, the overlaps between the maximal quantum mean values and the classical bound
  of the CHSH inequality, we show that the concurrence of
  a high-dimensional multipartite pure state can be exactly expressed by these overlaps. We further derive a lower
  bound of the concurrence for any quantum states, which is
   tight for pure states. The lower bound not only imposes restriction on the non-locality distributions among the
   pairwise qubit states, but also supplies a sufficient condition
    for distillation of bipartite entanglement. Effective criteria for detecting genuine tripartite entanglement and
    the lower bound of concurrence for genuine tripartite entanglement
     are also presented based on such non-localities.
\end{abstract}

\maketitle

\emph{Introduction.}---
Quantum entanglement has been one of the most remarkable resource in quantum theory.
Multipartite and high-dimensional quantum entanglement has become increasingly important for
quantum communication\cite{bechmann2000,cerf2002}.
Recently, a growing interest has been devoted to investigation of such kind of quantum resource
\cite{krenn2014,howland2016,martin2017,yamasaki2018,Ritz2019}.
In \cite{kraft2018} the authors have derived a
general theory to characterize those high-dimensional quantum states for which the correlations
cannot simply be simulated by low-dimensional systems.

The Bell inequalities\cite{bell}
 are of great
importance for understanding the conceptual foundations
of quantum theory as well as for investigating quantum
entanglement, as Bell inequalities can be violated by quantum
entangled states. One of the most important Bell
inequalities is the Clauser-Horne-Shimony-Holt (CHSH)
inequality\cite{chsh} for two-qubit systems. In \cite{horo1995} Horodeckis have presented the
necessary and sufficient condition of violating the CHSH inequality by an arbitrary
mixed two-qubit state. In \cite{wang,qin} we have discussed the trade-off relation of CHSH
violations for multipartite-qubit states based on the norms of Bloch vectors.

A similar question to \cite{kraft2018} is that can we simulate high-dimensional quantum entanglement
by the violations of CHSH inequalities for pairwise-qubit states in two-dimensional subspaces?
We present here a positive solution to this problem (see Fig. 1). For simplicity, we call a ``two-qubit" state,
obtained by projecting high dimensional $d_1\otimes d_2$ bipartite space to $2\otimes 2$ subspaces, a qubit pair in the following.

\begin{figure}[h]
\begin{center}
\includegraphics[height=5cm,width=7cm]{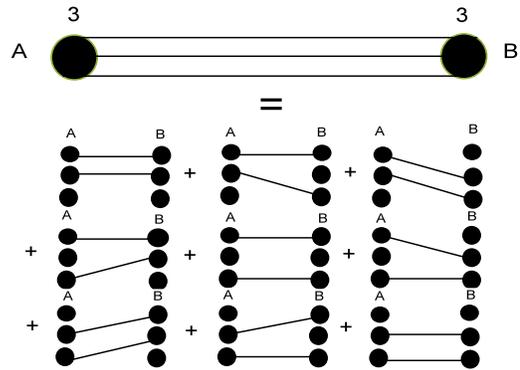}
\end{center}
\caption{The concurrence of any two-qutrit pure state is equal to the overlaps between the maximal
quantum mean values and the classical bound of the CHSH inequalities for nine pair of qubit states.
Thus entanglement can simply be simulated by the violation of CHSH inequalities of qubit pairs. The result holds for any pure states.}
\end{figure}

The second goal of this paper is to characterize genuine multipartite entanglement (GME)\cite{guhnerev} in high dimensional quantum systems.
As one of the
important type of entanglement, GME offers significant
advantage in quantum tasks comparing with bipartite entanglement
\cite{mule1}. In particular, it is the basic ingredient in
measurement-based quantum computation \cite{mule2}, and is
beneficial in various quantum communication protocols,
including secret sharing \cite{mule4,hillery}, extreme spin squeezing \cite{srensen}, high sensitivity in some
general metrology tasks \cite{toth}, quantum computing with
cluster states \cite{rauss}, and multiparty quantum network  \cite{mule3}.
Despite its significance, detecting and measuring such kind of entanglement turn out to be quite difficult.
To certify GME, an abundance of linear and nonlinear entanglement witnesses
\cite{huber2010,huber2014,vicente3,huber1,wu,sperling,eltschka,klockl,mark}, generalized
concurrence for genuine multipartite entanglement \cite{ma1,
ma2,gaot1,gaot2}, and Bell-like inequalities \cite{bellgme}, entanglement witnesses
were derived (see e.g. reviews \cite{guhnerev,siewert}) and a characterisation in terms of semi-definite
programs was developed \cite{jungnitsch,lancien}.
Nevertheless, the problem remains far from being satisfactorily solved.

In this paper we investigate entanglement by considering the overlap between the maximal quantum mean value
and the classical bound of the CHSH inequality. The overlap is used to derive a lower bound of concurrence
for any multipartite and high dimensional quantum states, which is tight for pure states. Thus we show that
the concurrence in any quantum systems can be equivalently represented by the violations of the CHSH inequalities
for qubit pairs. The lower bound not only imposes restriction on the non-locality distributions among qubit pairs,
 but also supplies a sufficient condition for bipartite distillation of entanglement. Criteria for detection genuine
 tripartite entanglement (GTE) and lower bound of GTE concurrence are further presented by the overlaps.
 We then show by examples that these criteria and the lower bound can detect more genuine tripartite entangled
 states than the existing criteria do.

We start with a short introduction of the generators of special
orthogonal group $SO(d)$ and the CHSH Bell inequalities. The
generators of $SO(d)$ can be introduced according to the
transition-projection operators $T_{st}=|s\rangle\langle t|,$ where
$|s\ra$, $s=1,\cdots,d$, are the orthonormal eigenstates of a linear
Hermitian operator on ${\mathcal H}_d$. Set $P_{st}=T_{st}-T_{ts},$
where $1\leq s < t \leq d$. We get a set of $\frac{d(d-1)}{2}$
operators that generate $SO(d)$. Such kind of operators(which will
be denoted by $L_{\alpha}, \alpha=1,2,\cdots,\frac{d(d-1)}{2}$) have
$d-2$ rows and $d-2$ columns with zero entries. For two-qubit
quantum systems, the CHSH Bell operators\cite{chsh} are defined by
\be\label{chsh} {I_{CHSH}}=A_1\otimes B_1+A_1\otimes B_2+A_2\otimes
B_1-A_2\otimes B_2, \ee where $A_i=\vec{a_i}\cdot
\vec{\sigma}_A=\sum\limits_{k=1}^3a_i^k\sigma_A^k$,
$B_j=\vec{b_j}\cdot
\vec{\sigma}_B=\sum\limits_{l=1}^3b_j^l\sigma_B^l$,
$\vec{a_i}=(a_i^1,a_i^2,a_i^3)$ and $\vec{b_j}=(b_j^1,b_j^2,b_j^3)$
are real unit vectors satisfying $|\vec{a_i}|=|\vec{b_j}|=1$,
$i,j=1,2$, $\sigma_{A/B}^{1,2,3}$ are Pauli matrices. The CHSH
inequality says that if there exist local hidden variable models to
describe the system, the inequality $ |\la{I_{CHSH}} \ra|\leq2 $
must hold. For any two-qubit state $\rho$, one defines the matrix
$X$ with entries $x_{kl}={\rm Tr}\{\rho\sigma_k\otimes\sigma_l\}$,
$k,l=1,2,3$. Horodeckis have computed in \cite{horo1995} the maximal
quantum mean value $\gamma=\max|\la I_{CHSH}
\ra_{\rho}|=2\sqrt{\tau_1+\tau_2}$, where the maximum is taken for
all the CHSH Bell operators $I_{CHSH}$ in Eq.(\ref{chsh}), $\tau_1,
\tau_2$ are the two greater eigenvalues of the matrix $X^tX,$ $X^t$
stands for the transposition of $X$.

\emph{Distribution of high-dimensional entanglement in qubit
pairs.}--- Let us first consider general $d\times d$ bipartite
quantum systems in vector space $\mathcal
{H}_{AB}=\mathcal{H}_A\otimes \mathcal{H}_B$ with dimensions
$dim\,{\mathcal {H}_A}=dim\,{\mathcal {H}_B}=d$, respectively.
Denote by $L_\alpha^A$ and $L_\beta^B$ the generators of special
orthogonal groups $SO(d)$. Let $\vec{a_i}$, $\vec{b_j}$ and
$\sigma_i$s denote unit vectors and Pauli matrices, respectively.
Set $\vec{\sigma}=(\sigma_1,\sigma_2,\sigma_3)$. We define the
operators $A_i^{\alpha}$ (resp. $B_j^{\beta}$) from $L_{\alpha}$
(resp. $L_{\beta}$) by replacing the four entries on the positions
of the nonzero 2 rows and 2 columns of $L_{\alpha}$ (resp.
$L_{\beta}$) with the corresponding four entries of the matrix
$\vec{a_i}\cdot\vec{\sigma}$ (resp. $\vec{b_j}\cdot\vec{\sigma}$),
and keeping the other entries of $A_i^{\alpha}$ (resp.
$B_j^{\beta}$) zero. We then define the following CHSH type Bell
operator:
\begin{eqnarray}\label{bellop}
\mathcal{B}_{\alpha\beta}=A^{\alpha}_1\otimes B^{\beta}_1+A^{\alpha}_1\otimes B^{\beta}_2
+A^{\alpha}_2\otimes B^{\beta}_1-A^{\alpha}_2\otimes B^{\beta}_2.
\end{eqnarray}

Set $y_{\alpha\beta}={\rm Tr}\{(L_\alpha^A)^{\dag}L_\alpha^A\otimes
(L_\alpha^B)^{\dag}L_\beta^B\rho\}$. If $y_{\alpha\beta}\neq 0$, we
define
$\rho_{\alpha\beta}=\frac{1}{y_{\alpha\beta}}L_\alpha^A\otimes
L_\alpha^B\rho(L_\alpha^A\otimes L_\alpha^B)^{\dag},
\gamma_{\alpha\beta}(\rho)=\frac{1}{y_{\alpha\beta}}\max {\rm
Tr}\{{\mathcal {B}}_{\alpha\beta}\rho\}$, where the maximum is taken
over all the Bell operators ${\mathcal {B}}_{\alpha\beta}$ of the
form(\ref{bellop}). Otherwise we set $\rho_{\alpha\beta}=0$ and
$\gamma_{\alpha\beta}(\rho)=0$.
We further define that
\begin{eqnarray}\label{defq}
\mathcal{Q}_{\alpha\beta}(\rho)=\max \{\gamma_{\alpha\beta}^2(\rho)-4,0\},
\end{eqnarray}
which will be called the CHSH overlaps of $\rho$.
If we can find a certain pair of $\alpha\beta$ such that $\mathcal{Q}_{\alpha\beta}(\rho)>0$, then the two qudit
state $\rho\in{\mathcal{H}_{AB}}$ must be nonlocal as a Bell inequality is violated.

For a bipartite pure state $\rho_{AB}=|\psi\ra\la\psi|\in
\mathcal{H}_{AB}$, the concurrence is defined by \cite{conc,anote}
$\mathcal {C}(|\psi\rangle)=\sqrt{2\left(1-{\rm
Tr}\rho_A^2\right)}$, where $\rho_A={\rm Tr}_B\rho_{AB}$ is the
reduced density matrix. For a mixed state
$\rho=\sum_{i}p_{i}|\psi_{i}\ra\la\psi_{i}|$, $p_{i}\geq0$,
$\sum_{i}p_{i}=1$, the concurrence is defined as the convex-roof:
$\mathcal{C}(\rho)=\min\sum_{i}p_{i}\mathcal{C}(|\psi_{i}\ra),$
minimized over all possible pure state decompositions.

We are ready to represent concurrence in high dimensional systems by the CHSH overlaps $\mathcal{Q}_{\alpha\beta}(|\psi\ra)$.
\begin{theorem}
For any two qudit pure quantum state $|\psi\ra\in{\mathcal
{H}}_{AB}$, we have
\be\label{thm1}
\mathcal{C}^2(|\psi\ra)=\frac{1}{4}\sum_{\alpha\beta}y^2_{\alpha\beta}
\mathcal{Q}_{\alpha\beta}(|\psi\ra).
\ee
\end{theorem}

\textbf{Proof.}
For any two-qubit pure state $|\phi\ra=\sum_{i,j=1}^2a_{ij}|ij\ra$,
the concurrence $\mathcal{C}(|\phi\ra)$ and $\mathcal{Q}_{\alpha\beta}(|\phi\ra)$ are preserved under any local
unitary operations. Thus to prove the theorem, we just need to consider the Schmidt decomposition of
$|\phi\ra=\sum_{i=1}^2\lambda_i|ii\ra,$ where $\sum_{i=1}^2\lambda_i^2=1$. One computes
$\mathcal {C}^2(|\phi\ra)=4\lambda_1^2\lambda_2^2,$
and
$\mathcal {Q}_{11}(|\phi\ra)=\frac{16\lambda_1^2\lambda_2^2}{(\lambda_1^2+\lambda_2^2)^2}.$
By $\sum_{i=1}^2\lambda_i^2=1$, we get
\be\label{aaq}\mathcal{C}^2(|\phi\ra)=\frac{1}{4}\mathcal{Q}_{11}(|\phi\ra).\ee

Then we consider two-qudit pure state $|\psi\ra=\sum_{i,j=1}^d a_{ij}|ij\ra$,
$\mathcal{C}^2(|\psi\ra)$ can be equivalently represented by \cite{anote,ou}
\begin{equation}\label{aa}
\mathcal
{C}^2(|\psi\rangle)=\sum_{\alpha\beta}|
{C}_{\alpha\beta}(|\psi\rangle\langle \psi|)|^2=4\sum_{i<j}^{d}\sum_{k<l}^{d}|a_{ik}a_{jl}-a_{il}a_{jk}|^2,
\end{equation}
where ${C}_{\alpha\beta}(|\psi\rangle\langle \psi|)=\langle\psi|\widetilde{\psi}_{\alpha\beta}\rangle$,
$|\widetilde{\psi}_{\alpha\beta}\rangle=(L_{\alpha}\otimes L_{\beta})|\psi^*\rangle$,
and $L_\alpha$ and $L_\beta$, $\alpha,\beta=1, ..., d(d-1)/2$, are
the generators of group $SO(d)$.
From (\ref{aaq}) and (\ref{aa}) we have
\begin{eqnarray*}
\mathcal{C}^2(|\psi\ra)&=&\sum_{\alpha\beta}y^2_{\alpha\beta}
\mathcal{C}^2(|\psi_{\alpha\beta}\ra)=\frac{1}{4}\sum_{\alpha\beta}y^2_{\alpha\beta}
\mathcal{Q}_{\alpha\beta}(|\psi\ra).
\end{eqnarray*}
\hfill \rule{1ex}{1ex}

It should be noted that in \cite{liang} the authors have computed the optimal expectation
value of the CHSH operator in \cite{Braunstein} for bipartite pure states in $d$ dimension.
The result in \cite{liang} is derived by representing the Hilbert space as a direct sum of two-dimensional subspaces,
plus a one-dimensional subspace if $d$ is odd.
While our Theorem 1 above shows that the concurrence of any bipartite high dimensional states can be equivalently
represented by the CHSH overlaps of qubit pairs.
We can further derive a lower bound for concurrence as an outgrowth of the Theorem.

\begin{theorem}
For any bipartite mixed qudit quantum state $\rho\in{\mathcal
{H}}_{AB}$, we have
\be\label{thm2}
\mathcal{C}(\rho)\geq\frac{1}{2}\sqrt{\sum_{\alpha\beta}y^2_{\alpha\beta}
\mathcal{Q}_{\alpha\beta}(\rho)}.
\ee
\end{theorem}

\textbf{Proof.} Assume that $\rho=\sum_i {p_i}|\psi_i\ra\la\psi_i|$,
$\sum p_i=1$, be the optimal ensemble decomposition such that
$\mathcal{C}(\rho) =\sum_ip_i\mathcal{C}( |\psi_i\ra)$. We have
\begin{eqnarray*}
\mathcal{C}(\rho)&=&\sum_ip_i\mathcal{C}( |\psi_i\ra)\ge\sqrt{\sum_{\alpha\beta}C^2(y_{\alpha\beta}\rho_{\alpha\beta})}\\
&=&\sqrt{\sum_{\alpha\beta}y^2_{\alpha\beta}\mathcal{C}^2(\rho_{\alpha\beta})}\\
&=&\sqrt{\sum_{\alpha\beta}y^2_{\alpha\beta}\sum_iq_i\mathcal{C}^2(\rho^i_{\alpha\beta})}\\
&=&\frac{1}{2}\sqrt{\sum_{\alpha\beta}y^2_{\alpha\beta}
\sum_iq_i\mathcal{Q}_{\alpha\beta}(\rho^i_{\alpha\beta})}\\
&\ge&\frac{1}{2}\sqrt{\sum_{\alpha\beta}y^2_{\alpha\beta}
\mathcal{Q}_{\alpha\beta}(\rho)},
\end{eqnarray*}
where in the first inequality we have used the theorem 1 given in
\cite{ou}. \hfill \rule{1ex}{1ex}

In \cite{ou} the authors have derived a lower bound of concurrence
in terms of the concurrence of $2\times 2$-dimensional substates.
Here we present a lower bound of concurrence in terms of the CHSH
overlaps. Theorems above can be directly generalized to multipartite
case. An $N$-partite pure state in $\mathcal{H}_1\otimes
\mathcal{H}_2\otimes\cdots\otimes \mathcal{H}_N$ is generally of the
form, \be\label{mpure} |\Psi\ra=\sum\limits_{i_{1},i_{2},\cdots
i_{N}=1}^{d}a_{i_{1}i_{2}\cdots i_{N}}|i_{1}i_{2}\cdots i_{N}\ra,
\ee where $a_{i_{1}i_{2}\cdots i_{N}}$s are entries of a complex
vector with unit length. Let $\alpha$ and $\alpha^{'}$ (resp.$\beta$
and $\beta^{'}$) be subsets of the subindices of $a$, associated to
the same sub Hilbert spaces but with different summing indices.
$\alpha$ (or $\alpha^{'}$) and $\beta$ (or $\beta^{'}$) span the
whole space of the given sub-indix of $a$. The generalized
concurrence of $|\Psi\ra$ is then given by \cite{anote},
\begin{eqnarray}\label{defmc}
\mathcal{C}_{d}^{N}(|\Psi\ra)=\sqrt{\sum\limits_{p}
\sum\limits_{\{\alpha,\alpha^{'},\beta,\beta^{'}\}}^{d}
|a_{\alpha\beta}a_{\alpha^{'}\beta^{'}}-a_{\alpha\beta^{'}}a_{\alpha^{'}\beta}|^{2}},
\end{eqnarray}
where $\sum\limits_{p}$ stands for the summation over all possible
combinations of the indices of $\alpha$ and $\beta$. In
(\ref{defmc}) we have ignored a overall constant factor for
simplicity. For a mixed state
$\rho=\sum_{i}p_{i}|\psi_{i}\ra\la\psi_{i}|$, $p_{i}\geq0$,
$\sum_{i}p_{i}=1$, the concurrence is defined by the convex-roof:
\begin{eqnarray}\label{conrho}
\mathcal{C}_{d}^{N}(\rho)=\min\sum_{i}p_{i}\mathcal{C}_{d}^{N}(|\psi_{i}\ra),
\end{eqnarray}
minimized over all possible pure state decompositions.

By using Theorem 1 and Eq.(\ref{defmc}) we obtain for any
$N$-partite pure state in the form of (\ref{mpure}) that
\be\label{meq}
(\mathcal{C}_{d}^{N})^2(|\Psi\ra)=\frac{1}{4}\sum_p\sum_{\alpha\beta}(y^p_{\alpha\beta})^2
\mathcal{Q}^p_{\alpha\beta}(|\Psi\ra), \ee where $y^p_{\alpha\beta}$
and $\mathcal{Q}^p_{\alpha\beta}(|\Psi\ra)$ are defined similarly to
the bipartite case by considering $|\Psi\ra$ as a bipartite state
with respect to partition $p$.

For any $N$-partite mixed state $\rho_N$, we get \be\label{lbm}\mathcal{C}_{d}^{N}(\rho_N)
\geq\frac{1}{2}\sqrt{\sum_p\sum_{\alpha\beta}(y^p_{\alpha\beta})^2
\mathcal{Q}^p_{\alpha\beta}(\rho_N)},\ee
where $y^p_{\alpha\beta}$ and $\mathcal{Q}^p_{\alpha\beta}(\rho_N)$ are defined similarly to the bipartite
case by considering $\rho_N$ as a bipartite state with respect to partition $p$.

Eq.(\ref{lbm}) will be tight if $\rho_N$ is an N-partite pure state.
Thus we conclude that the concurrence of any high-dimensional multipartite pure states can be equivalently
represented by the CHSH overlaps of a series of pairwise-qubit states (See Fig. 2 for three-qubit systems as an example).

\begin{figure}[h]
\begin{center}
\includegraphics[height=5cm,width=7cm]{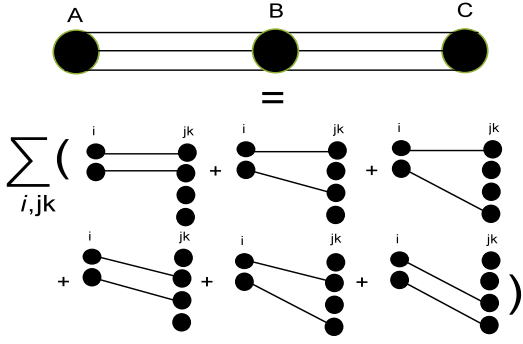}
\end{center}
\caption{The concurrence of any three-qubit pure state is given by the CHSH overlaps of six pairs of qubit states.}
\end{figure}

\emph{Detection and measure of genuine tripartite entanglement by the CHSH overlaps.}---
In this section we consider tripartite quantum systems
$\mathcal{H}_{123}=\mathcal{H}_{1}\otimes\mathcal{H}_{2}\otimes\mathcal{H}_{3}$, with $\dim \mathcal{H}_{i}=d, i=1,2,3.$

 A tripartite state $\rho  \in \mathcal{H}_{123}$ can be expressed as $\rho  = \sum {{p_\alpha }} \left| {{\psi _\alpha }}
 \right\rangle \left\langle {{\psi_\alpha }} \right|$, where $0<p_\alpha\leq 1$,
$\sum {{p_\alpha }}  = 1$, $\left| {{\psi _\alpha }} \right\rangle \in \mathcal{H}_{123}$ are normalized pure states.
If all $\left| {{\psi _\alpha }} \right\rangle$ are biseparable, namely, either
$\left| {{\psi _\alpha }} \right\rangle = \left| {\varphi _\alpha ^1} \right\rangle  \otimes \left| {\varphi _\alpha ^{23}}
\right\rangle $ or $\left| {{\psi _\beta }} \right\rangle  = \left| {\varphi _\beta ^2} \right\rangle
\otimes \left|{\varphi _\beta ^{13}} \right\rangle $ or $\left| {{\psi _\gamma }} \right\rangle
= \left| {\varphi _\gamma ^3} \right\rangle  \otimes \left|{\varphi _\gamma ^{12}} \right\rangle$,
where $\left| {\varphi_\gamma^i} \right\rangle$ and $\left| {\varphi_\gamma^{ij}} \right\rangle$
denote pure states in $\mathcal{H}_i^d$ and $\mathcal{H}_i^d \otimes \mathcal{H}_j^d$ respectively,
then $\rho $ is said to be bipartite separable. Otherwise, $\rho $ is called genuine tripartite entangled.

For any $\rho\in\mathcal{H}_{123}$, we define $X=\max_{\alpha\beta} \mathcal{Q}_{\alpha\beta}^{1|23},
Y=\max_{\alpha\beta} \mathcal{Q}_{\alpha\beta}^{2|13}$
and $Z=\max_{\alpha\beta} \mathcal{Q}_{\alpha\beta}^{3|12}$.

\begin{theorem}\label{thm4}
For any pure tripartite state $|\psi\ra$,
$\min\{X, Y, Z\}>0$ holds if and only if $|\psi\ra$ is genuine tripartite entangled.
\end{theorem}

\textbf{Proof.} According to the definition, any bi-separable pure state $|\psi\ra$ must be either
$\left| {\psi} \right\rangle = \left| {\varphi^1} \right\rangle  \otimes \left| {\varphi ^{23}} \right\rangle $
 or $\left| {{\psi}} \right\rangle  = \left| {\varphi^2} \right\rangle  \otimes \left|{\varphi^{13}} \right\rangle $
  or $\left| {{\psi}} \right\rangle  = \left| {\varphi^3} \right\rangle  \otimes \left|{\varphi^{12}} \right\rangle$.
  On the contrary, if $|\psi\ra$ is GTE (not bi-separable), then it must be not in any bi-separable form, which can be
  represented by violating all the CHSH inequalities for any qubit pairs of $|\psi\ra$. This can be further represented
  by $\min\{X, Y, Z\}>0$ according to the definition of $X, Y,$ and $Z$. \hfill \rule{1ex}{1ex}

The sufficient and necessary condition for detecting GTE in Theorem 3 can be generalized to any pure multipartite
 quantum states. In the following we derive a sufficient condition to detect GTE for any tripartite mixed quantum states.

\begin{theorem}\label{thm5}
If $\rho\in{\mathcal
{H}}_{123}$ is bipartite separable, then
\be\label{th6}
X+Y+Z\leq 8
\ee always holds. Thus if (\ref{th6}) is violated, then $\rho$ is of GTE.
\end{theorem}

\textbf{Proof.}
For any bipartite separable pure state, say, $\left| {\psi} \right\rangle = \left| {\varphi^1} \right\rangle
 \otimes \left| {\varphi ^{23}} \right\rangle$, one gets $X=0, Y\leq 4$ and $Z\leq 4$, which proves (\ref{th6}).

Now consider a mixed bipartite separable state with ensemble decomposition $\rho  = \sum p_{\alpha} \left|
 \psi _\alpha  \right\rangle \left\langle \psi_\alpha \right|$ with $\sum p_\alpha= 1$. By noticing that all
  $X, Y$ and $Z$ are convex function of $\rho$ and the summation of convex functions is still a convex function, we have
\be
X+Y+Z\leq \sum_\alpha p_\alpha(X_\alpha+Y_\alpha+Z_\alpha)\leq 8\sum_\alpha p_\alpha = 8.
\ee
\hfill
\rule{1ex}{1ex}

The GTE concurrence for tripartite quantum systems defined below is proved to be a well defined measure\cite{ma1,ma2}.
For a pure state $|\psi\ra\in \mathcal{H}_{123}$, the GTE concurrence is defined by
\begin{eqnarray*}
\mathcal{C}_{GTE}(|\psi\ra)=\sqrt{\min\{1-{\rm Tr}(\rho_1^2),1-{\rm Tr}(\rho_2^2),1-{\rm Tr}(\rho_3^2)\}},
\end{eqnarray*}
where $\rho_i$ is the reduced matrix for the $i$th subsystem.
For mixed state $\rho\in \mathcal{H}_{123}$, the GTE concurrence is then defined by the convex roof
\begin{eqnarray}
\mathcal{C}_{GTE}(\rho)=\min\sum_{\{p_{\alpha},|\psi_{\alpha}\ra\}}
p_{\alpha}\mathcal{C}_{GTE}(|\psi_{\alpha}\ra).
\end{eqnarray}
The minimum is taken over all pure ensemble decompositions of $\rho$.
Since one has to find the optimal ensemble for the minimization, the GTE concurrence is hard to compute.
In the following we present a lower bound of GTE concurrence in terms of
$\mathcal{Q}_{\alpha\beta}s$.

\begin{theorem}\label{thm7}
Let $\rho  \in {\mathcal{H}_{123}}$ be a tripartite qudits quantum state.
Then one has
\be\label{gmelb}
C_{GTE}(\rho)\geq \frac{1}{6\sqrt{2}}\sum_p\sqrt{\sum_{\alpha\beta}(y_{\alpha\beta}^p)^2
\mathcal{Q}_{\alpha\beta}^p(\rho)}-\frac{2}{3}\sqrt{\frac{d-1}{d}},
\ee
where the partitions $p\in\{1|23, 2|13, 3|12\}$.
\end{theorem}

\textbf{Proof.} We start the proof with a pure state. Let
$\rho=|\psi\ra\la\psi|\in {\mathcal{H}_{123}}$ be a pure quantum
state. From the result in Theorem 1, we have
\begin{eqnarray*}
\sqrt{1-tr\rho_1^2}=
\frac{1}{2\sqrt{2}}(\sum_{\alpha\beta}(y^{1|23}_{\alpha\beta})^2
\mathcal{Q}^{1|23}_{\alpha\beta}(|\psi\ra))^{\frac{1}{2}}
\end{eqnarray*}
and
\begin{eqnarray*}
\sqrt{1-tr\rho_k^2}\leq \sqrt{\frac{d-1}{d}},~~ k=2,3.
\end{eqnarray*}
Therefore,
\begin{eqnarray*}
\sqrt{1-tr\rho_1^2}\geq
\frac{1}{6\sqrt{2}}\sum_p\sqrt{\sum_{\alpha\beta}(y_{\alpha\beta}^p)^2
\mathcal{Q}_{\alpha\beta}^p(\rho)}-\frac{2}{3}\sqrt{\frac{d-1}{d}}.
\end{eqnarray*}
Similarly, we get
\begin{eqnarray*}
\sqrt{1-tr\rho_k^2}\geq
\frac{1}{6\sqrt{2}}\sum_p\sqrt{\sum_{\alpha\beta}
(y_{\alpha\beta}^p)^2
\mathcal{Q}_{\alpha\beta}^p(\rho)}-\frac{2}{3}\sqrt{\frac{d-1}{d}},
\end{eqnarray*}
where $k=2,3.$ Then according to the definition of GME concurrence,
we derive \be\label{22} C_{GTE}(|\psi\ra)\geq
\frac{1}{6\sqrt{2}}\sum_p\sqrt{\sum_{\alpha\beta}(y_{\alpha\beta}^p)^2
\mathcal{Q}_{\alpha\beta}^p(\rho)}-\frac{2}{3}\sqrt{\frac{d-1}{d}}.
\ee

Now we consider a mixed state $\rho\in {\mathcal{H}_{123}}$ with the
optimal ensemble decomposition
$\rho=\sum_{x}q_{x}|\psi_{x}\ra\la\psi_{x}|$, $\sum_{x}q_{x}=1$,
such that the GTE concurrence attains its minimum. By (\ref{22}) one
gets
\begin{eqnarray*}
&&C_{GTE}(\rho)=\sum_{x}q_{x}C_{GME}(|\psi_{x}\ra)\\
&\geq&
\frac{1}{6\sqrt{2}}\sum_{p,x}q_x\sqrt{\sum_{\alpha\beta}(y_{\alpha\beta}^p(|\psi_{x}\ra))^2
\mathcal{Q}_{\alpha\beta}^p(|\psi_{x}\ra)}-\frac{2}{3}\sqrt{\frac{d-1}{d}}\\
&\geq&\frac{1}{6\sqrt{2}}\sum_p\sqrt{\sum_{\alpha\beta}(y_{\alpha\beta}^p)^2
\mathcal{Q}_{\alpha\beta}^p(\rho)}-\frac{2}{3}\sqrt{\frac{d-1}{d}},
\end{eqnarray*}
where we have used $\sum_{x}q_{x}=1$ and inequality
$\sum_i\sqrt{\sum_jx^2_{ij}}\geq\sqrt{\sum_j(\sum_ix_{ij})^2}$.
\hfill \rule{1ex}{1ex}

Let us now  consider an example to illustrate further the significance of our result for detection of GTE.

\textbf{Example 1:} Consider the quantum state $\rho\in
\mathcal{H}_1^d \otimes \mathcal{H}_2^d\otimes \mathcal{H}_3^d$,
\be\label{sigma} \sigma(x)=x|\psi\ra\la\psi|+\frac{1-x}{d^2}I, \ee
where $|\psi\ra=\frac{1}{\sqrt{d}}\sum\limits_{i=1}^d|iii\ra$ and
$I$ stands for the identity operator.

By the positivity of $X+Y+Z-8$, we get the ranges of $x$ for different $d$ such that $\sigma(x)$ is GTE (see table I).
\begin{widetext}
\begin{center}
\begin{table}[ht]
\centering
\caption{Detection of GTE of $\sigma(x)$ by Theorem 4 (Range 1), Theorem 5 (Range 2), Theorem in \cite{GEli2015}
(Range 3), Theorem 1 in \cite{vicente3, klockl} (Range 4).}
\begin{tabular}{lccc}
\hline
\textbf{Dimension} &~\textbf{d=2}~ &~\textbf{d=3} ~&~\textbf{d=4} ~\\ \hline
Range 1   &~~$x>0.839708~~$ &~~$x>0.699544$~~ &~~$x>0.567035~~$ \\ \hline
Range 2   &$x>0.788793$ &$x>0.731621$ &$x>0.705508$ \\ \hline
Range 3   &$x>0.8532$       &$x>0.83485$   &$x>0.82729$ \\ \hline
Range 4   &$x>0.87$     &$x>0.89443$  &$x>0.91287$ \\ \hline
\end{tabular}
\end{table}
\end{center}
\end{widetext}
The data in Table I show that Theorem 4 and 5 in this letter, independently, detect more genuine tripartite
entangled states than that in \cite{GEli2015}(by the lower bound of multipartite concurrence), \cite{vicente3}
and in \cite{klockl}(by the correlation tensor norms).

\emph{The CHSH overlaps and distillation of entanglement.}--- The
CHSH overlaps defined in (\ref{defq}) can be also applied to
distillation of entanglement. In \cite{dur2001} D\"{u}r has shown
that there exist some multi-qubit bound entangled (non-distillable)
states that violate a Bell inequality. Ac\'{\i}n further proves in
\cite{acin2002} that for all states violating this inequality there
is at least one splitting of the parties into two groups such that
some pure state entanglement can be distilled under this partition.
The relation between violation of Bell inequalities and bipartite
distillability of multi-qubit states is further studied in
\cite{lee2009}. The lower bound (\ref{lbm}) has also a close
relationship with bipartite distillation of any multipartite and
high dimensional states. Note that a density matrix $\rho$ is
distillable if and only if there are some projectors $A,B$ that map
high-dimensional spaces to two-dimensional ones and a certain number
$n$ such that the state $A\otimes B\rho^{\otimes n}A\otimes B$ is
entangled\cite{disiff}. Thus if \be
\label{dism}\max_{\alpha\beta}\mathcal{Q}^p_{\alpha\beta}(\rho^{\otimes
n})>0 \ee for a certain partition $p$, then there exists one
submatrix of matrix $\rho^{\otimes n}$, which is entangled in a
$2\times 2$ space. Hence we get that $\rho$ is bipartite distillable
in terms of bipartition $p$. The constraint (\ref{dism}) is
equivalent to the strict positivity of the lower bound in
(\ref{lbm}). Note that
$\max_{\alpha\beta}\mathcal{Q}^p_{\alpha\beta}(\rho^{\otimes n})$ is
generally not an invariant under local unitary operations on the
state $\rho$. It is helpful to select proper local unitary
operations to enhance the value of
$\max_{\alpha\beta}\mathcal{Q}^p_{\alpha\beta}(\rho^{\otimes n})$
from 0 to a positive number. Since the separability is kept
invariant under local unitary operations, we have that if
$\max_{U_1,U_2,\cdots,U_n}\max_{\alpha\beta}\mathcal{Q}_{\alpha\beta}(U_1\otimes
U_2\otimes \cdots\otimes U_n\rho^n U_1^{\dag}\otimes
U_2^{\dag}\otimes \cdots\otimes U_n^{\dag})>0$ hold for proper
unitary $U_i$s, ${i=1,..,n}$, then $\rho$ is entangled and bipartite
distillable.

\textbf{Example 2:} Consider the quantum state $\rho\in {\rm H}_1^d
\otimes {\rm H}_2^d$, \be
\rho(x)=x|\psi\ra\la\psi|+\frac{1-x}{d^2}I, \ee where
$|\psi\ra=\frac{1}{\sqrt{d}}\sum\limits_{i=1}^d|ii\ra$ and $I$
stands for the identity operator.

By the positivity of
$\max_{\alpha\beta}\mathcal{Q}_{\alpha\beta}(\rho)$, one computes
the ranges of $x$ for different $d$ such that $\rho$ is non-local
and 1-distillable (see table I, Range 1). Range 2 is derived by the
reduction criterion (RC), as violation of RC is a sufficient
condition of entanglement distillation\cite{rc2002,cerf1999}.

\begin{widetext}
\begin{center}
\begin{table}[ht]
\centering \caption{Distillation of non-locality and entanglement
for $\rho(x)$ in Example 2:}
\begin{tabular}{lcccccc}
\hline \textbf{Dimension} &\textbf{d=2} &\textbf{d=3} &\textbf{d=4}
&\textbf{d=5}
   &\textbf{d=6} &\textbf{d=7}\\ \hline
Range 1   &$x>0.707107$ &$x>0.616781$ &$x>0.546918$ &$x>0.491272$
&$x>0.445903$ &$x>0.408205$\\ \hline Range 2   &$x>0.33333$
&$x>0.25$ &$x>0.2$ &$x>0.16667$ &$x>0.142857$ &$x>0.125$\\ \hline
\end{tabular}
\end{table}
\end{center}
\end{widetext}

\textbf{Example 3:}  Consider the quantum state $\rho\in {\rm H}_1^d
\otimes {\rm H}_2^d\otimes {\rm H}_3^d$, \be\label{sigma}
\sigma(x)=x|\psi\ra\la\psi|+\frac{1-x}{d^2}I, \ee where
$|\psi\ra=\frac{1}{\sqrt{d}}\sum\limits_{i=1}^d|iii\ra$ and $I$
stands for the identity operator.

To check the bipartite 1-distillability of $\sigma(x)$, we compare
$\max_{\alpha\beta}\mathcal{Q}^p_{\alpha\beta}(\rho)$ with $0$ for
$p=1|23, 2|13,$ and $3|12$. One computes the ranges of $x$ for
different $d$ such that $\rho$ is 1-distillable (see table II, Range
1).

\begin{widetext}
\begin{center}
\begin{table}[ht]
\centering \caption{Bipartite 1-distillation of entanglement for
$\sigma(x)$ in Example 3:}
\begin{tabular}{lcccc}
\hline \textbf{Dimension} &\textbf{d=2} \ \ \ \ \ &\textbf{d=3}
&\textbf{d=4} \ \ \ \ \  &\textbf{d=5}\\ \hline Range   &$x>0.54692$
\ \ \ \ \  &$x>0.34917$\ \ \ \  &$x>0.23182$ \ \ \ \ \  &$x>0.16188$
\\ \hline
\end{tabular}
\end{table}
\end{center}
\end{widetext}

\emph{Conclusions and remarks.}--- In summary we have considered the
CHSH overlaps for quantum states. It has been shown that the
concurrence of any multipartite and high dimensional pure states can
be equivalently represented by the CHSH overlaps of a series of
``two-qubit" states. Based on the overlaps sufficient condition for
distillation of entanglement have been obtained. As another
application of the CHSH overlaps, we have further presented criteria
for detecting GME and lower bound of GME concurrence for tripartite
quantum systems. For tripartite pure states, a sufficient and
necessary condition is derived to detect GME, while for tripartite
mixed states, we have obtained effective sufficient conditions and
lower bounds for GME concurrence. An important question that needs
further discussion is to find a criterion that discriminates W state
and GHZ state.

Recently high dimensional bipartite systems like in NMR and nitrogen-vacancy defect center
have been successfully used in quantum computation and simulation experiments\cite{du}.
Our results present a plausible way to measure the multipartite concurrence in these systems and
to investigate the roles played by the multipartite concurrence in these quantum information processing. Our approach
 of the CHSH overlaps of qubit pairs can also be employed to investigate the distributions of other quantum correlations
  in high dimensional systems. Another important question that needs
further discussion is to find a criterion that discriminates W state
and GHZ state, as GTE is a common property of W state and GHZ state,
but there is no local unitary transformation to relate them.

\noindent{\bf Acknowledgments}\, \, This work is supported by the NSFC No. 11775306, 11701568, 11701128 and 11675113;
the Fundamental Research Funds for the Central Universities Grants No.17CX02033A, 18CX02023A and 19CX02050A; the Shandong
Provincial Natural Science Foundation No. ZR2016AQ06, ZR2017BA019, and Key Project of Beijing Municipal Commission of
Education under No. KZ201810028042.

\end{document}